\documentclass[preprint,12pt]{elsarticle}

\usepackage[utf8]{inputenc}
\usepackage{amssymb}
\usepackage{graphicx}%
\usepackage{multirow}%
\usepackage{amsmath,amssymb,amsfonts}%
\usepackage{amsthm}%
\usepackage{mathrsfs}%
\usepackage[title]{appendix}%
\usepackage{xcolor}%
\usepackage{textcomp}%
\usepackage{manyfoot}%
\usepackage{booktabs}%
\usepackage{algorithm}%
\usepackage{algorithmicx}%
\usepackage{algpseudocode}%
\usepackage{listings}%
\usepackage{subcaption}%
\usepackage{multirow}%
\usepackage{float}%
\usepackage{makecell}
\usepackage{tabularx}
\usepackage{array}
\usepackage{placeins}%
\usepackage{float}%
\usepackage{geometry}
\usepackage{longtable}
\usepackage{graphicx}
\usepackage{placeins}
 \usepackage[utf8]{inputenc} 
\usepackage{adjustbox} 
\usepackage{comment}  %
\newcolumntype{L}[1]{>{\raggedright\arraybackslash}p{#1}}

\journal{}

\begin{document}

\begin{frontmatter}



\title{Machine Learning-Driven Crystal System Prediction for Perovskites Using Augmented X-ray Diffraction Data}

\author[inst1]{Ansu Mathew\corref{cor1}}
\author[inst1]{Ahmer A. B. Baloch}
\author[inst1]{Alamin Yakasai}
\author[inst1]{Hemant Mittal}
\author[inst1]{Vivian Alberts}
\author[inst1]{Jayakumar V. Karunamurthy}

\affiliation[inst1]{organization={Research and Development Center, Dubai Electricity and Water Authority},  
            addressline={MBR Solar Park},  
            city={Dubai},  
            country={United Arab Emirates}
           }
\cortext[cor1]{Corresponding authors: Ansu Mathew, Email: \texttt{ansu.mathew@dewa.gov.ae}}  

\begin{abstract}
Prediction of crystal system from X-ray diffraction (XRD) spectra is a critical task in materials science, particularly for perovskite materials which are known for their diverse applications in photovoltaics, optoelectronics, and catalysis. In this study, we present a machine learning (ML)-driven framework that leverages advanced models, including Time Series Forest (TSF), Random Forest (RF), Extreme Gradient Boosting (XGBoost), Recurrent Neural Network (RNN), Long Short-Term Memory (LSTM), Gated Recurrent Unit (GRU), and a simple feedforward neural network (NN), to accurately classify crystal systems, point groups, and space groups from XRD data of perovskite materials. The TSF model outperformed other approaches, demonstrating the advantage of treating XRD data as time-series data, effectively capturing spectral dynamics and sequential dependencies inherent in diffraction patterns. To address class imbalance and enhance model robustness, we integrated feature augmentation strategies such as Synthetic Minority Over-sampling Technique(SMOTE), class weighting, jittering, and spectrum shifting, along with efficient data preprocessing pipelines. TSF model with SMOTE augmentation achieved outstanding performance for crystal system prediction, with a mathew's correlation coefficient(MCC) of 0.9, an F1 score of 0.92, and an accuracy of 97.76\%. For point group prediction, the model achieved an MCC of 0.79, an F1 score of 0.83, and a balanced accuracy of 95.27\%. For space group prediction, the TSF model with class weighting and jittering achieved an MCC of 0.8, an F1 score of 0.84, and a balanced accuracy of 95.18\%. The model demonstrated exceptional performance for classes with distinct symmetry features, such as cubic crystal system, point groups like $3m$ and $m-3m$, and space groups like $Pnma$ and $Pnnn$, where precision, recall, F1 score, and MCC values consistently approached 1.00. This work underscores the potential of ML in XRD data analysis, enabling faster, more precise characterization of complex materials like perovskites.By streamlining material identification, this approach will accelerate the discovery of novel materials for advanced technologies.
\end{abstract}




\end{frontmatter}



\section{Introduction}\label{sec:sample1}
{
X-ray diffraction (XRD) is an indispensable technique in materials science, extensively utilized for the discovery and characterization of novel materials~\cite{alghadeer2024machine}. By directing a beam of X-rays onto a sample, typically in powder form, XRD exploits the principle of constructive and destructive interference arising from the interaction of X-rays with the atomic planes of a crystalline structure. The resulting diffraction patterns, recorded as intensity peaks at specific angles, provide critical insights into the sample's crystallographic structure, including lattice parameters, phase composition, and crystalline quality~\cite{Szymanski2024}. This technique has been instrumental in elucidating the fundamental properties and behaviours of crystalline materials, making it a cornerstone in materials characterization.Despite its broad application, the interpretation of XRD data to deduce crystal systems has traditionally been time-consuming and dependent on expert judgment. However, the integration of m achine learning (ML) has significantly transformed this domain~\cite{dong2022deepxrd}. ML algorithms have demonstrated strong capabilities in automating XRD analysis, supporting both high-throughput symmetry classification and the inference of structure-property relationships~\cite{massuyeau2022perovskite}. For perovskites in particular—where polymorphism, low symmetry, and structural distortions are common—accurate symmetry classification is essential for analyzing electronic, optical, and stability-related behavior~\cite{alsaui2022highly}. A growing body of work has focused on ML-based crystal system and space group classification from XRD data. Among them, CrystalMELA has established itself as a leading platform for ML-driven classification of polycrystalline materials~\cite{Corriero2023,Rizzi2024,Settembre2022}. Their models and web-based interface enable accessible crystal system prediction using diverse powder diffraction datasets. These efforts have been complemented by recent work in interpretable AI for chemistry and drug discovery from the same group~\cite{Lomuscio2025,Delre2023}, which demonstrates how explainable ML can guide practical decision-making in molecular design. In parallel, Simon Billinge’s group has contributed a suite of ML techniques for diffraction-based structure solution. Their research spans ab initio structure determination from noisy or nanocrystalline XRD data using diffusion models~\cite{Guo2025}, interpretable multimodal analysis of XANES and PDF data~\cite{Narong2025}, and ML-driven structure mining directly from pair distribution functions~\cite{Kjaer2024,Terban2021}. These contributions have broadened the scope of ML in crystallography beyond classification, addressing inverse problems and embracing data imperfection.}

\begin{table}[htbp]
\centering
\caption{{Summary of literature on machine learning applied to XRD-based structure prediction.}}
\label{tab:literature_summary}
\scriptsize
\resizebox{\textwidth}{!}{
{
\begin{tabular}{|p{3.1cm}|p{1.5cm}|p{2.3cm}|p{2.3cm}|p{2.5cm}|p{1.8cm}|}
\hline
\textbf{Reference} & \textbf{No. of Samples} & \textbf{Material Family} & \textbf{ML Algorithm} & \textbf{Target} & \textbf{Performance} \\
\hline
Kaufmann et al. & 50,000 & ICSD & CNN (ResNet50, Xception) & Bravais Lattice & 91--93\% \\
\hline
Dong et al. & 4270 (ABC3), 37211 (Ternary) & Perovskites & DeepXRD & Spectrum prediction & Pearson r: 0.68 \\
\hline
Suzuki et al. & 188,607 & ICSD & Random Forest & Space group & 83.62\% \\
\hline
X. Zhao et al. & Unknown & Crystalline & DNN & crystal system (Synchrotron) & $\sim$90\% \\
\hline
Greasley \& Hosein & -- & Biomedical & SVM, CNB, ANN, KNN & Multi-phase quantification & High phase fidelity \\
\hline
Chen et al. & 60,000+ & Inorganic & ResNet (CRNN) & crystal system (100 types) & High generalizability \\
\hline
Massuyeau et al. & 23 & Hybrid Perovskites & CNN & Structure classification & 92\% \\
\hline
Marchenko et al. & 100 & Perovskites & DT, RF, XGB, CatBoost & Topology, Dimensionality & 82--86\% \\
\hline
B. Do Lee et al. & 15,000 & MP + ICSD & FCN, CGCNN, T-encoder & Symmetry, Properties & 92.12\% (Crystal System) \\
\hline
J. W. Lee et al. & 1,785,405 & Inorganic & CNN & Phase ID, Quantification & ID: $\sim$100\%, Quant: 86\% \\
\hline
Oviedo et al. & 115 & Halide Perovskites & All-CNN + Augmentation & Dimensionality, Space group & 93\%, 89\% \\
\hline
Salgado et al. & 204,654 & ICSD & CNN & Space group, Crystal system & 86\% \\
\hline
Yanxon et al. & 294,400 & Nickel, Battery & RF, GBM, KNN, SVM & Artifact detection & TP $>$ 95\%, TN $>$ 99.9\% \\
\hline
B. Zhao et al. & 401 & Explosives & SVM & Noise detection & $\sim$50\% \\
\hline
\end{tabular}
}
}
\end{table}

{Table~\ref{tab:literature_summary} provides a comparative overview of recent machine learning models applied to XRD-based structural classification, highlighting their datasets, algorithms, targets, and achieved performance across material domains.} Integration of ML into XRD data analysis has contributed significantly to the advancement of material characterization, enhancing both efficiency and accuracy. Kaufmann et al. \cite{kaufmann2020crystal} tackled a critical limitation in electron backscatter diffraction (EBSD), which traditionally relied on expert input and Hough-based matching for phase identification. 50,000  patterns from nine distinct materials were used for the CNN model, with the Bravais lattice correctly classified at 93.5\\\% accuracy using ResNet50 and 91.2\% using Xception. In the domain of biomedical materials, Greasley and Hosein demonstrated the utility of support vector machines (SVMs), complement naive Bayes (CNB), and artificial neural networks (ANNs) for multi-phase identification for biomedical applications using  XRD spectra. These models performed comparably to advanced techniques like k-nearest neighbors (KNN), effectively quantifying phase fractions and showcasing the applicability of conventional ML methods in automating complex material analysis tasks \cite{greasley2023exploring}. The application of ML to perovskite materials has also seen substantial progress. Dong et al. \cite{dong2022deepxrd}addressed the challenge of predicting XRD spectra directly from perovskite material composition with their DeepXRD model.  Different loss functions for XRD similarity measures were evaluated on ABC3-XRD (4,270 samples) and Ternary-XRD (37,211 samples), showing that Pearson correlation achieved the highest peak position match percentage (0.681 for ABC3-XRD and 0.678 for Ternary-XRD), followed by Cosine similarity (0.673, 0.667) and MSE (0.626, 0.612).  Massuyeau et al. \cite{massuyeau2022perovskite} developed a CNN to classify hybrid lead halide perovskite structures using a limited dataset of 23 samples. Achieving 92\% accuracy, the model effectively identified structural features such as unit-cell volumes and Pb-ion spacings, demonstrating ML's potential in distinguishing perovskites from non-perovskites and accelerating the discovery of hybrid materials. Chen et al. \cite{Chen2024}enhanced automated crystal system determination by employing a convolutional residual neural network (ResNet) trained on over 60,000 samples representing 100 structural types. Their model significantly reduced reliance on expert knowledge and enabled the incorporation of new structural categories without retraining, demonstrating its adaptability and utility in material discovery. Suzuki etal \cite{Suzuki2018} utilized random forest models to predict crystal system from XRD peak positions. With an accuracy of 83.62\% using 188,607 samples from the ICSD, their approach demonstrated the computational efficiency of random forest regression over deep learning, making it an effective tool for high-throughput crystallography. Zhao et al. \cite{Zhao2023} explored automated XRD data analysis for large-scale synchrotron experiments using deep neural networks (DNNs). Their study highlighted the critical role of labeled experimental data in improving model accuracy, achieving over 90\% performance when combining synthetic and experimental datasets for XRD mapping in liquid-phase experiments. {
Recent studies have begun to explore ML approaches for classifying perovskite structures from XRD patterns, but most remain limited to general datasets or focus on coarse-level classification (e.g., crystal system only). For example, Marchenko et al.\ \cite{marchenko2025machine} applied a decision tree classifier to perovskite XRD data and reported accuracy of 82--86\% for topology and dimensionality classification. However, many such models are not robust to experimental artifacts like texture or overlapping peaks. Furthermore, few models target full-stack symmetry prediction (crystal system $\rightarrow$ point group $\rightarrow$ space group), which is essential for understanding symmetry-governed phenomena in device performance.} 
Lee et al. \cite{Lee2022}developed a CNN model trained on powder XRD patterns from Materials Project (MP) and ICSD datasets. Achieving 92.12\% accuracy in crystal system classification, their approach combined high-dimensional data with latent space embeddings, revealing well-separated clusters that linked symmetry to material properties. Oviedo et al. \cite{Oviedo2019}tackled the limitations of small XRD datasets by implementing a physics-informed data augmentation strategy. Their CNN model achieved 93\% accuracy for dimensionality classification and 89\% for space group classification, providing interpretability through global mean pooling-derived class maps, thus facilitating experimental feedback. Szymanski et al. \cite{Szymanski2023} developed adaptive ML-driven XRD systems for real-time phase identification during in situ experiments. Their system dynamically adjusted predictions based on early diffraction data, accelerating phase detection and demonstrating ML's potential in autonomous material characterization. Salgado et al. \cite{Salgado2023}created a deep learning model to classify space groups and crystal systems with 86\% accuracy using ICSD samples. By optimizing Bragg's Law and decision-making processes, their model efficiently adapted to experimental datasets, enhancing its practical applicability. Yanxon et al. \cite{Yanxon2023}applied ML models, including gradient boosting and random forests, to identify artifacts in XRD data for nickel and battery materials. With a true positive rate exceeding 95\% and a true negative rate of 99.9\%, their work highlighted ML's role in improving data quality and reliability.

crystal system identification using XRD spectra is crucial for understanding phase purity, lattice parameters, and defects in perovskite materials, which directly influence their optoelectronic properties. Structural stability, determined through XRD analysis, plays a key role in the long-term performance and degradation resistance of perovskite solar cells. The ability to classify crystal system from XRD spectra using machine learning provides a rapid, automated, and scalable approach to analyzing experimental diffraction data. Traditionally, phase identification requires expert interpretation and comparison with databases, which can be time-consuming and subjective.{
The primary bottlenecks in structure prediction from XRD patterns of perovskite materials arise not just from structural complexity but also from experimental artifacts and data quality limitations. Thin-film perovskites often exhibit strong preferred orientation (texture), low signal-to-noise ratios, and missing reflections due to limited angular ranges in measurement setups. These factors obscure symmetry-relevant features in the diffraction pattern and make conventional phase identification techniques (e.g., Rietveld refinement or database matching) error-prone or inconclusive—especially for distinguishing between closely related space groups or detecting distortions in pseudo-cubic structures.
}.  {In the context of solar energy, halide perovskites such as MAPbI$_3$, FAPbI$_3$, and CsPbI$_3$ are among the most promising materials due to their high absorption coefficients, tunable band gaps, and defect tolerance. The symmetry of these materials affects key performance parameters such as charge transport anisotropy, exciton binding energy, and degradation pathways. For example, the cubic phase often corresponds to optimal optoelectronic performance, whereas deviations from high symmetry (e.g., due to strain or processing conditions) may lead to undesirable phase segregation or trap-assisted recombination. Thus, fast and accurate symmetry classification from minimal XRD data is essential for both material screening and device optimization workflows.
}

{
To address these limitations, our study presents a symmetry classification framework designed specifically for perovskite materials. We use a time-series representation of XRD data and apply Time Series Forest (TSF) models, which are inherently more interpretable and capable of capturing localized features across the diffraction signal. This modeling approach avoids reliance on either image-based transformations or handcrafted peak features, making it better suited for real-world, noisy, and imbalanced perovskite datasets. We also implement targeted data augmentation strategies (SMOTE, jittering) to handle class imbalance and benchmark our model against deep learning alternatives. This architecture enables accurate classification of under-represented symmetries, which are critical for photovoltaic materials research but often ignored in large-scale ML datasets. Our method enables high-throughput screening of perovskite materials, assisting in experimental workflows by reducing manual effort and accelerating materials discovery. This is particularly useful in large-scale combinatorial studies where numerous compositions are synthesized, and automated phase identification can significantly streamline the characterization process. Additionally, the framework can be integrated into autonomous experimental platforms, where real-time classification of XRD data informs synthesis optimization and guides iterative design. Our work employs state-of-the-art ML models to enhance the prediction of crystal system from XRD spectra for perovskite materials. By incorporating advanced feature augmentation strategies such as time-series decomposition and engineered physical descriptors based on crystallographic principles, our approach ensures a comprehensive representation of the diffraction data. The workflow begins with data pre-processing, including interpolation and normalization, followed by data augmentation techniques such as jittering, class weighting, SMOTE, peak scaling, and spectrum shifting to enhance data diversity and address class imbalances. Machine learning models, including Time Series Forest, Random Forest, XGBoost, and Neural Networks, are then trained on the augmented dataset to classify crystal system, point groups, and space groups (symbols). Among these, Time Series Forest demonstrated superior performance, particularly when combined with data augmentation strategies like SMOTE and jittering. This integrated framework combines robust pre-processing, augmentation, and model optimization to enable accurate, scalable, and efficient crystal system prediction, ultimately advancing high-throughput characterization for complex materials like perovskites.
}

\section{Methodology}\label{sec2}
\begin{figure}[h!]
    \centering
    \includegraphics[width=\columnwidth]{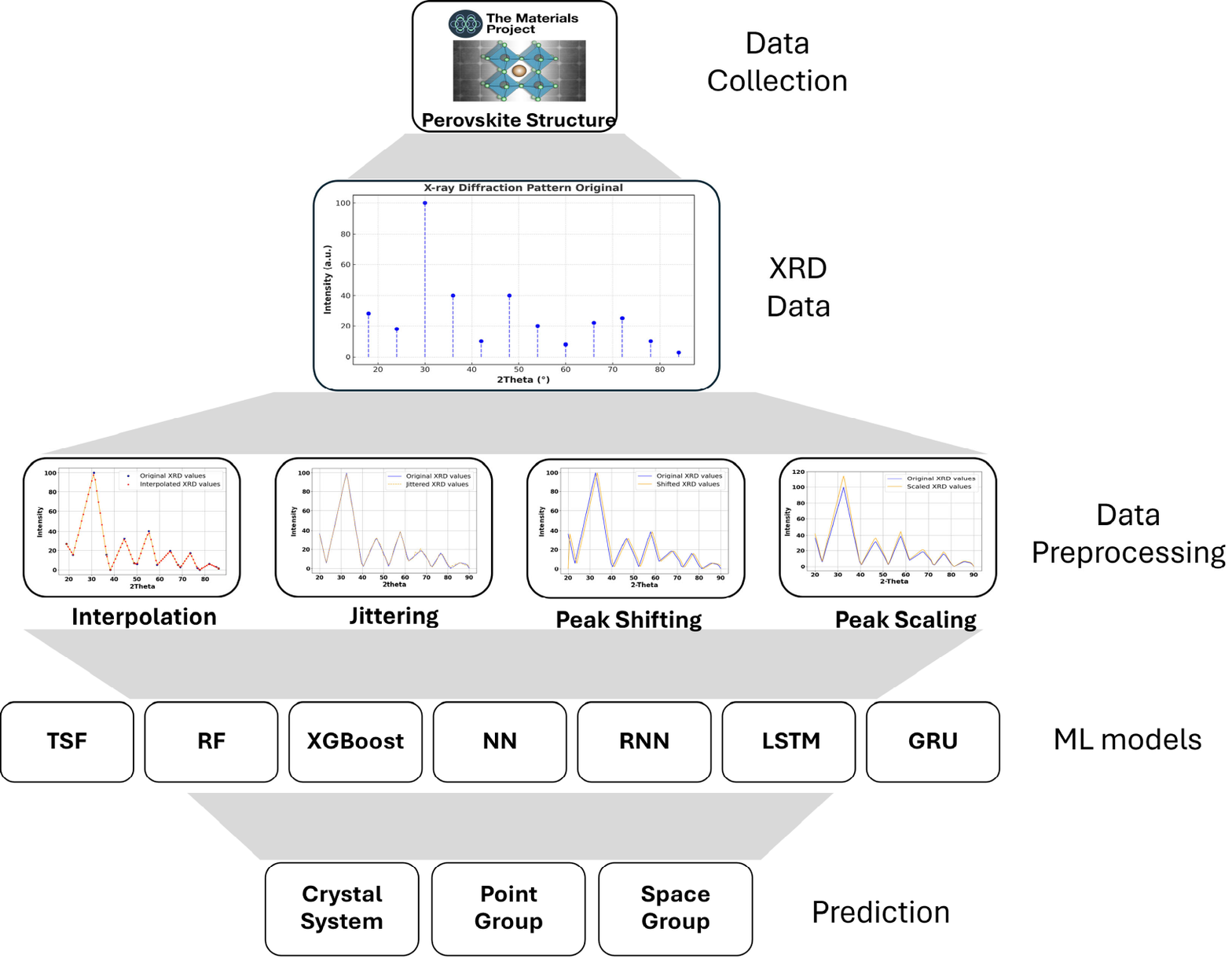}
    \caption{ML pipeline for crystallographic classification, from data collection and preprocessing to feature augmentation and model training, predicting crystal system, point group, and space group}
    \label{fig:methodolgy}
\end{figure}

XRD classification using machine learning involves several steps, including data collection, preprocessing, data augmentation, model selection, and prediction of crystal system, point group, and space group. The methodology is illustrated in Figure\ref{fig:methodolgy} and explained below.
\subsection{Data Collection and Dataset Description}\label{subsec2}
\subsubsection{Source of Data and Details.}\label{subsubsec2}

Materials Project \cite{jain2013commentary} repository was used to create the XRD dataset. It provides extensive computational tools and databases to analyze XRD patterns, helping in material identification and characterization. We accessed the data via the repository's API using Pymatgen (Python Materials Genomics), an open-source Python library for materials analysis \cite{ong2013python}. We  used robocrystallographer to generate textual descriptions of crystal system to identify 9,284 perovskite materials. After removing the duplicates and merging the initial set, the process yielded 4,009 unique perovskite Material IDs. We extracted the XRD data and the crystal system details of each material using the pymatgen library. The datasets, XRD and associated crystal system, were combined into a single data frame using Python for training ML models. {
During dataset curation, Robocrystallographer was used to generate structured textual descriptions summarizing octahedral connectivity, dimensionality, and coordination environments. These qualitative descriptors enabled screening for canonical ABX$_3$-type perovskite topologies and helped ensure structural consistency across the dataset. While not used directly as model inputs, they played a critical role in filtering invalid or atypical frameworks. Symmetry labels including crystal system, point group, and space group were derived using the \texttt{SpacegroupAnalyzer} class in the \texttt{pymatgen} library, which interfaces with \texttt{spglib}. Symmetry determination was applied to DFT-relaxed geometries using a standard tolerance (\texttt{symprec} = 0.1 \AA), ensuring physical validity and robustness to small distortions. This automated pipeline avoids heuristic thresholds and guarantees reproducibility in symmetry classification.
}. In this process, it was important to examine the basic aspects of crystal symmetry. A crystal system is a way to group crystals based on the shape and size of their unit cells, including the lengths of the edges and the angles between them. There are seven types of crystal systems: cubic, tetragonal, orthorhombic, hexagonal, trigonal, monoclinic, and triclinic. A point group describes the symmetry elements present in a crystal, such as rotational and reflectional symmetries, without considering translational symmetry. There are 32 unique point groups, each defining how a crystal system remains unchanged under specific symmetry operations. A space group extends the concept of point groups by incorporating translational symmetry, fully defining the three-dimensional symmetry of a crystal lattice. There are 230 unique space groups, which account for the complete symmetry of crystalline materials, including screw axes and glide planes. Comprehending these classifications facilitates the interpretation of structural characteristics and contributes to improving the accuracy of machine learning models in materials prediction. Figure \ref{fig:histogram} the histogram plot of the crystal system, point group, and space group along with a donut chart illustrating the contribution of each class to the overall dataset. These plots provide an overall viewpoint of the dataset extracted from the Material Project repository.

\subsubsection{Characteristics of the XRD Data
}
XRD data obtained from the Material Project repository vary in length. All XRD data should be the same length to feed the data into the ML model. We performed data interpolation (Figure \ref{fig:augmentation}a) to standardize the data length,  resulting in each sample having a fixed input length of 90 for the ML models.\color{black}

The peak intensities of the XRD data depict the scattering power of atoms in the crystal lattice. The scattering power of an atom is determined by its atomic number, the number of electrons in its outermost shell, and the spatial distribution of those electrons within the atom. Therefore, rather than taking the raw XRD data, we  also considered features of XRD data for model training. The features we considered are:
\begin{itemize}
    \item The five highest peak intensities: \( p_1, p_2, p_3, p_4, p_5 \)
    \item The corresponding diffraction angles: \( \theta_1, \theta_2, \theta_3, \theta_4, \theta_5 \)
    \item Weighted peak intensities: \( p_1\theta_1, p_2\theta_2, \dots, p_5\theta_5 \)
    \item Skewness of the intensity distribution
\end{itemize}
We also explored using the wavelet transform and the Fourier transform of the XRD signal, instead of using the raw XRD signal directly. We observed that using raw XRD data yields better performance than using engineered features. The results are presented in Table \ref{tab:Features}.in the supplementary information.\color{black}
\label{subsubsec2}
\subsubsection{Data Preprocessing}\label{subsubsec2}
In the dataset, we observed that a few classes of the point group and space group have a very low number of samples. We performed data selection to ensure a more representative distribution. The histogram plot shown in Figure \ref{fig:histogram} represents the class distribution in decreasing order of sample count, with the dashed line indicating the cutoff point for selecting samples for point group and space group. We selected the initial 15 classes for both the point group and space group. For the point group, we selected classes up to '2', where the number of samples was 23. For the space group, we selected up to 'Pc', where the number of samples was 30.

{
To maintain a one-to-one correspondence between each XRD pattern and a unique symmetry label, we removed polymorphic duplicates—i.e., entries with identical chemical formulas but different space or point groups. This decision avoids ambiguity in class labels and ensures that the classification task is well-posed for supervised learning. In future work, we aim to extend the framework to explicitly incorporate polymorphism and phase transition pathways by exploring multi-label, probabilistic, or temperature-aware classification models. This will enhance the applicability of ML-based XRD analysis to experimental datasets where such complexities are unavoidable.
}.Data normalization is one of the pre-processing steps before feeding the data into the ML model. This ensure that input features are on a similar scale, which can be crucial for training the ML model. However, for XRD data, normalization is not necessary because the data represents intensities corresponding to the diffraction angle, and its the characteristics closely resemble time- series data. We trained the machine learningML model in both scenarios: with and without normalization. Standard scaler normalization was applied by subtracting each value from the feature mean and dividing by the standard deviation. Additionally, we experimented with row-wise normalization, dividing each intensity value by the corresponding XRD’s highest value. We observed that the ML model performed better without XRD data normalization and the the related results are included the Table \ref{tab:row-norm} and \ref{tab:std-norm} of the  supplementary information.

\begin{figure}[htbp]
    \centering
    \includegraphics[height=0.9\textheight]{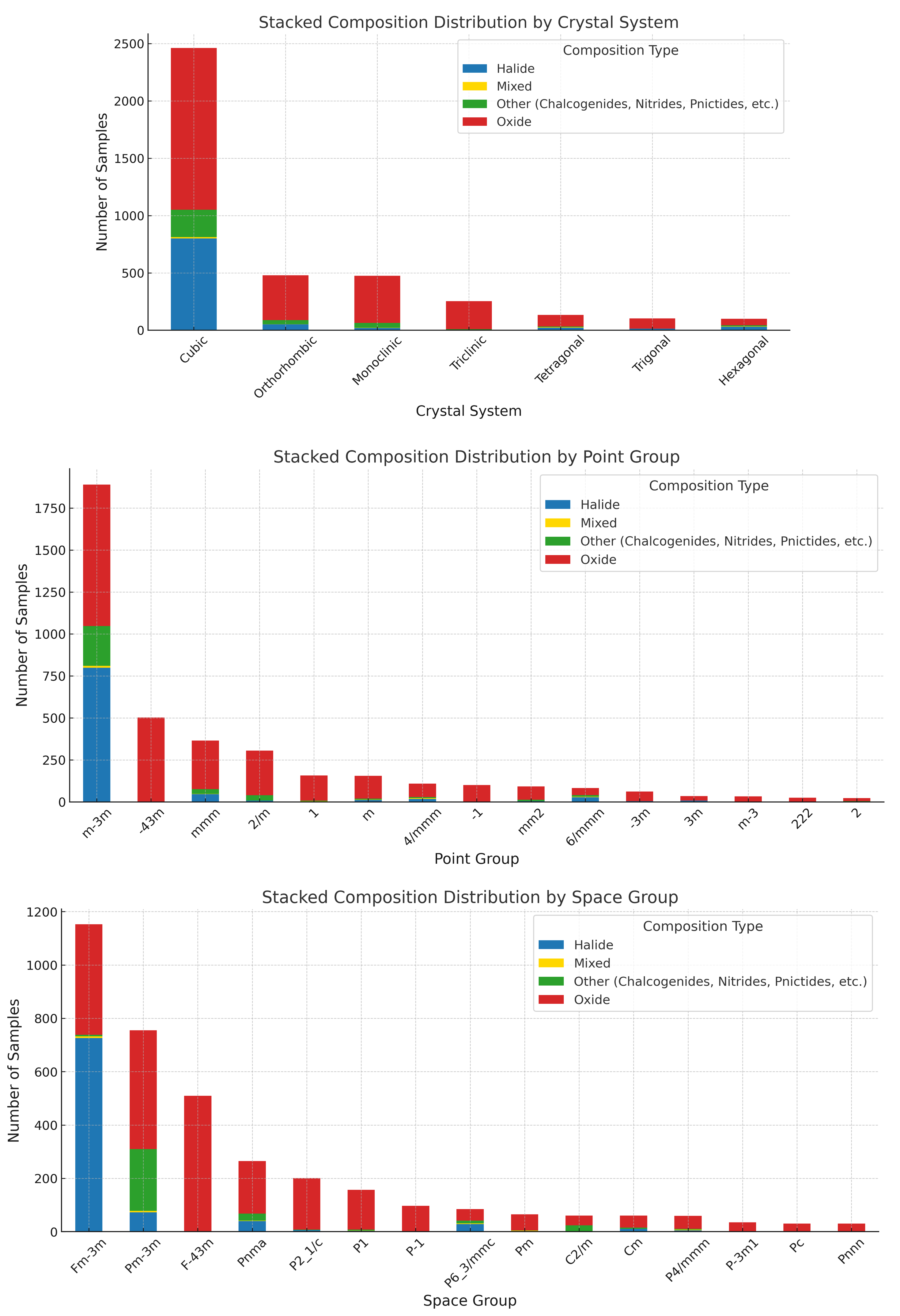}
    \caption{{Distribution of perovskite composition types: Oxide, Halide, Mixed, and Other—across the top 15 symmetry classes. }}
    \label{fig:composition}
\end{figure}

{
Perovskite materials exhibit a wide range of chemical compositions that influence their crystallographic symmetry and, consequently, the interpretability of their XRD patterns. To investigate this effect, we categorized the 4,009 perovskite entries into four composition classes based on the presence of anionic species: Oxide, Halide, Mixed (Halide + Oxide), and Other (Chalcogenides, Nitrides, Pnictides, etc.). This classification was determined by parsing chemical formulas and identifying key anionic groups such as O$^{2-}$, F$^{-}$, Cl$^{-}$, Br$^{-}$, I$^{-}$, S$^{2-}$, N$^{3-}$, and P$^{3-}$. The oxide group dominates the dataset (2,702 entries), followed by halides (945), with mixed and other classes contributing 25 and 336 entries, respectively.

{
To better understand the chemical diversity of our dataset and its influence on crystallographic symmetry, we performed a systematic compositional classification of all 4,009 perovskite entries extracted from the Materials Project. Each entry includes a chemical formula, space group, point group, and crystal system. Based on the anionic species present in the formula, we classified the dataset into four mutually exclusive categories: Oxide, Halide, Mixed Halide–Oxide, and Other.

\textbf{Oxide perovskites} (2,702 entries) contain only oxygen as the anion and no halide species. Representative compounds such as BaTiO$_3$, LaFeO$_3$, and YAlO$_3$ are known for their structural complexity, polymorphism, and high-temperature stability. These materials are distributed across a wide symmetry range including monoclinic, orthorhombic, tetragonal, and even triclinic systems. Their XRD patterns often feature peak broadening and multiplicity due to octahedral distortions and B-site cation displacements.

\textbf{Halide perovskites} (945 entries) include only halide anions (F$^-$, Cl$^-$, Br$^-$, I$^-$) and no oxygen. Examples like CsPbI$_3$, MAPbBr$_3$, and Rb$_2$TlBr$_6$ are typically crystallized in high-symmetry cubic or pseudocubic phases such as Fm$\bar{3}$m and Pm$\bar{3}$m. Their sharp and distinct diffraction peaks make them highly favorable for machine learning-based symmetry classification, especially in point groups like m$\bar{3}$m and $\bar{4}$3m.

\textbf{Mixed halide–oxide perovskites} (25 entries), such as K$_2$VOCl$_5$ and Ba$_2$YIO$_6$, contain both oxygen and at least one halide. These compounds show a broader structural variability, frequently adopting monoclinic or orthorhombic symmetry. Their limited representation in public databases and compositional complexity restrict meaningful statistical conclusions.

\textbf{Other compositions} (336 entries) consist of anionic species other than oxides or halides, including chalcogenides (e.g., S$^{2-}$, Se$^{2-}$), pnictides (e.g., N$^{3-}$, P$^{3-}$), and related species. Compounds such as K$_2$TiS$_4$, LaBiN$_3$, and Ba$_2$Sb$_2$Se$_7$ expand the chemical scope of the dataset. However, their unique bonding environments and low symmetry typically result in more complex XRD features and pose challenges for reliable model predictions.

We further analyzed how these composition classes map onto crystallographic symmetry labels. Halide perovskites are concentrated in high-symmetry crystal systems (notably cubic), space groups (Fm$\bar{3}$m, Pm$\bar{3}$m, F$\bar{4}$3m), and point groups (m$\bar{3}$m, $\bar{4}$3m), consistent with their symmetric ionic frameworks and minimal structural distortion. In contrast, oxides span a wider range of symmetry groups, including low-symmetry systems like monoclinic (P2$_1$/c), orthorhombic (Pnma), and triclinic (P$\bar{1}$), reflecting their structural flexibility and common distortions due to octahedral tilting and Jahn–Teller effects. ``Other'' compounds predominantly fall into monoclinic and orthorhombic categories, underscoring the structural diversity introduced by non-oxide, non-halide anions.
}

\begin{figure}[htbp]
    \centering
    \includegraphics[ width=\textwidth]{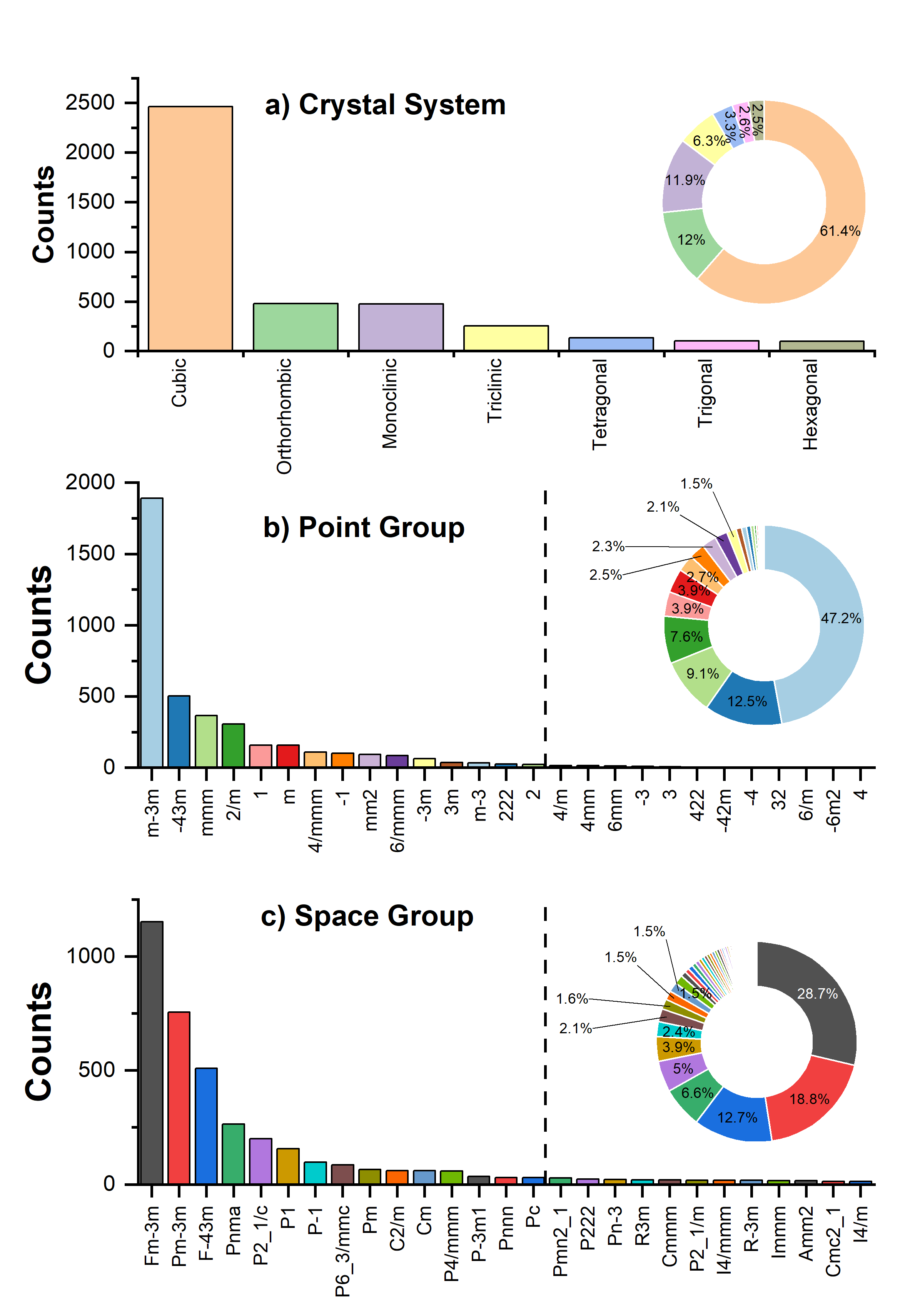} 
    \caption{Data distribution of crystal system, point group, and space group: (a) Histogram plot of the crystal system with a pie chart on the left showing the percentage contribution of each class; (b) Histogram plot of the point group, where the dashed line separates the classes chosen from the entire dataset. Classes up to point group 2, which has 23 values, were selected; (c) Histogram plot of the space group, where the initial 15 classes up to space group 'Pc', with a value count of 30, were considered for ML model development. }
    \label{fig:histogram}
\end{figure}

    
    

\begin{figure}[htbp]
    \centering
    \includegraphics[ width=\textwidth]{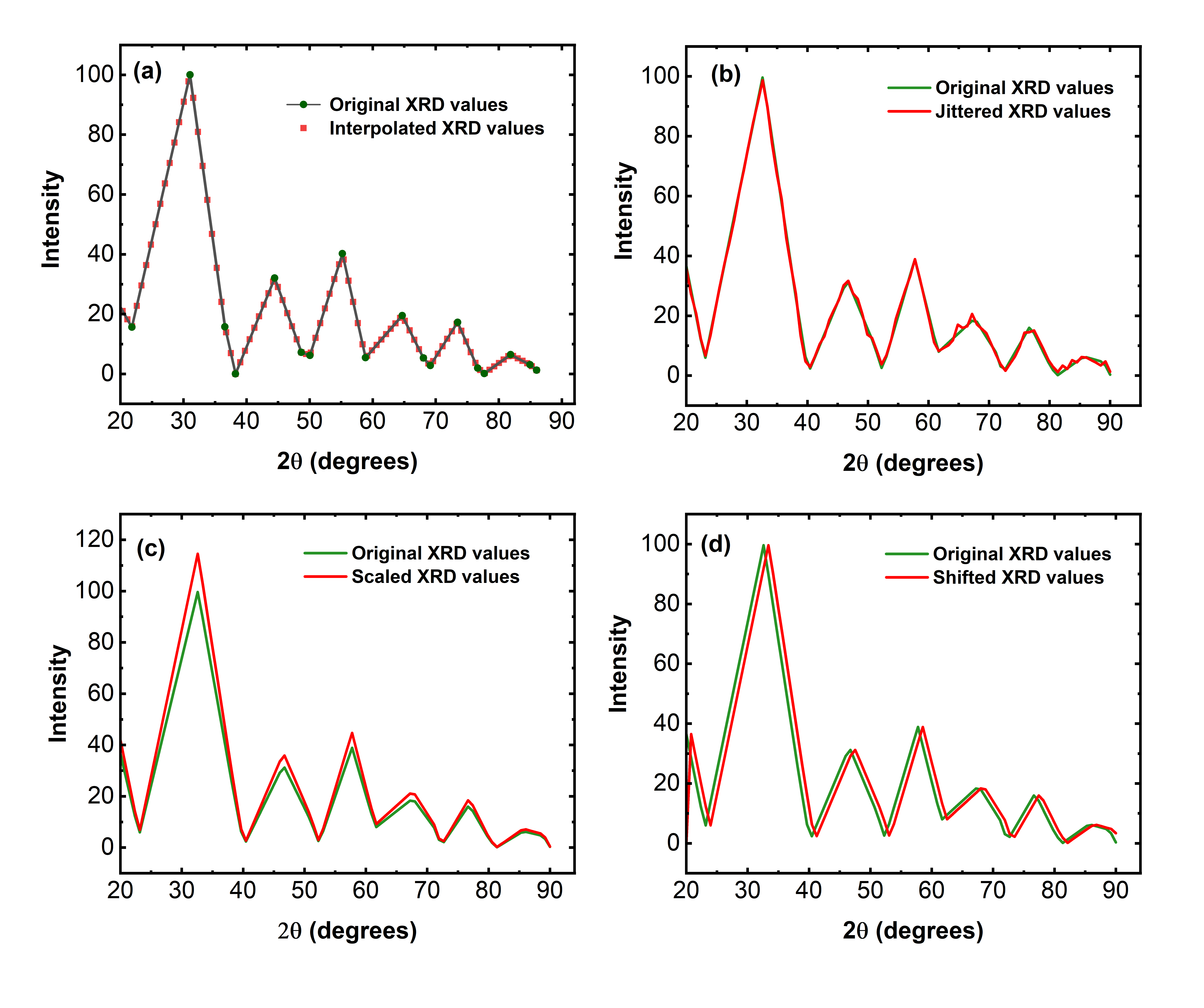} 
    \caption{ Data preprocessing Techniques for XRD patterns. Green line represents the original XRD data and red line represents augmented data(a) Interpolation: Comparison of original XRD values with interpolated values. (b) Jittering: Visualization of original and jittered XRD values. (c) Scaling: Comparison of original and scaled XRD values. (d) Shifting: Visualization of original and shifted XRD values. Each method demonstrates the effect of data augmentation on XRD patterns, preserving the overall intensity and peak positions.}
    \label{fig:augmentation}
\end{figure}

\subsection{{Data Filters}}\label{subsubsec2}

{
All structures used in this study were extracted from the Materials Project (MP) database using the Pymatgen API. A multi-stage filtering pipeline was applied to ensure data quality and physical relevance. First, perovskite-type structural motifs were identified. This was achieved using robocrystallographer descriptors. To enforce thermodynamic viability, only structures with negative formation energy per atom were retained, using the \texttt{formation\_energy\_per\_atom} field provided by MP. Furthermore, entries with an energy above the convex hull greater than 0.05 eV/atom were excluded, consistent with widely accepted thresholds in high-throughput DFT studies to indicate likely experimental synthesizability.

After this stability-based screening, deduplication was performed by retaining only one structure per unique combination of chemical composition and space group. This step removed polymorphs with differing symmetries to ensure label clarity for supervised learning tasks, yielding a pool of 4009 entries. Finally, to reduce class imbalance in the classification of point groups and space groups, we retained only the 15 most frequent classes in each case, applying a minimum threshold of 23 samples for point groups and 30 samples for space groups. This filtering protocol resulted in a clean, physically grounded, and statistically robust dataset for machine learning model development.
}
\subsection{Data Augmentation}\label{subsec2}
Data augmentation is a ML technique used to increase the size of dataset through different transformations. These transformations will help to create more diversity in the dataset, leading to the development of a more robust ML model. Increasing the data size leads to the improved model generalization due to the diverse datasets. In imbalanced dataset scenarios, we can use augmentation techniques to increase the datasets in the minority classes. We performed different XRD augmentations, like peak scaling (Figure \ref{fig:augmentation}c) and shifting(Figure \ref{fig:augmentation}d) \cite{Oviedo2019} . The metrics obtained after scaling and shifting were poor as shown in the supplementary material(Table \ref{tab:scaling} and \ref{tab:shifting}). While performing peak scaling, we lose the relative relationship between amplitudes, and shifting the XRD data leads to a loss of spectral information, making the augmented data unrealistic. In this work, we also used a technique called jittering for data augmentation. In jittering, we introduce a slight random noise to each signal value as shown in (Figure \ref{fig:augmentation}b). This random value is drawn from a normal distribution with a mean of zero and standard deviation. Excessive jittering may distort the signal. We carefully selected the value of $\sigma$ so that it would not affect the information in the signal; the $\sigma$ is 1. Synthetic Minority Over-sampling Technique(SMOTE) is another data augmentation technique specifically designed for dealing with imbalanced datasets. It oversamples the minority class by generating synthetic samples. This technique randomly samples from the minority samples identify the K nearest neighbours of these samples and generates synthetic samples by interpolating among the features of nearest neighbours.

\subsection{Machine Learning Models}\label{subsec2}
We explored various machine learning models and deep learning models such as Random forest, Timeseries Forest, XGBoost, neural networks. Additionally, we experimented with deep learning architectures such as RNN, LSTM, and GRU to evaluate their ability to capture sequential patterns within the XRD data. The data was divided into training and test sets using an 80:20 split, with stratification applied to preserve the class distribution within the target label.\color{black}We hyperparameter tuned each model using GridSearchCV, a systematic method to identify the optimum parameters. In GridSearchCV, we performed a 3-fold split of the data to ensure the robust performance of the model on different datasets. For deep learning models, we used KerasTuner for hyperparameter tuning—exploring the number of layers, units per layer, dropout rates, optimizers, and learning rates along with early stopping and model checkpointing to avoid overfitting and ensure optimal performance.\color{black}

\subsection{Evaluation Metrics}\label{subsec2}
We can describe the metrics as true positives, false positives and false negatives. True positives occur when the model correctly predicts positive samples. False negatives arise when the model incorrectly predicts true samples. True negatives occur when the model correctly predicts negative samples. False positives appear when the model incorrectly predicts a false sample as true.
Accuracy measures the overall correctness of the model. It is calculated using the formula represented in the equation \eqref{eq:accuracy} below. Accuracy is the ratio between the sum of true positives and true negatives to the whole number of instances. However, it does not account for the imbalanced nature of the datasets. Therefore, when we analyzed class-wise metrics, we considered binary accuracy, which is computed on a per-class basis by treating each class as a one-vs-rest binary problem. This helps evaluate how well the model distinguishes a specific class from all others, and it is particularly useful in imbalanced datasets.\color{black}
Precision(equation\eqref{eq:precision}) is the ratio of true positives to the sum of true positives and false positives. If the cost of predicting false positives is high, precision aids us in not making any incorrect positive predictions. Recall(equation\eqref{eq:recall}) is the ratio of True positives to the sum of true positives and false negatives. This measure is essential when capturing all true positives where the cost of missing a true positive is high. F1 Score (equation\eqref{eq:f1score}) is the harmonic mean of precision and recall, which ranges from 0 to 1. One indicates that the model achieved perfect precision and recall. Mathew's correlation coefficient(equation\eqref{eq:mcc}) covers all four parameters: TP, FP, TN, and FN. It implies that it takes care of the imbalanced dataset. The value of MCC ranges from -1 to +1. -1 means that the model completely mismatches between predictions and true labels; +1 corresponds to the best model; and 0 implies the model is not better than a random prediction.
\begin{equation}
    \text{Precision} = \frac{TP}{TP + FP}
    \label{eq:precision}
\end{equation}
\begin{equation}
    \text{Recall} = \frac{TP}{TP + FN}
    \label{eq:recall}
\end{equation}
\begin{equation}
    \text{F1 Score} = 2 \times \frac{\text{Precision} \times \text{Recall}}{\text{Precision} + \text{Recall}}
    \label{eq:f1score}
\end{equation}
\begin{equation}
    \text{Accuracy} = \frac{TP + TN}{TP + TN + FP + FN}
    \label{eq:accuracy}
\end{equation}
\begin{equation}
    \text{MCC} = \frac{TP \times TN - FP \times FN}{\sqrt{(TP + FP) \times (TP + FN) \times (TN + FP) \times (TN + FN)}}
    \label{eq:mcc}
\end{equation}

\subsection{{Limitations and Future Work}}\label{subsec2}

{
One of the persistent challenges in classifying symmetry information—particularly at the point and space group level—is the significant spectral overlap observed in XRD patterns across structurally similar classes. Monoclinic and orthorhombic systems, or space groups within the same Laue class, can exhibit nearly indistinguishable peak positions and intensities. This makes it difficult for machine learning models to resolve fine-grained distinctions, especially for low-symmetry structures with fewer unique reflections.

In our current study, we limited the input strictly to 1D powder XRD patterns to evaluate the classification capability using spectral data alone. To mitigate class overlap to some extent, we employed data augmentation techniques such as SMOTE and jittering, and applied class filtering to ensure balanced representation. However, we acknowledge that a more robust solution will likely require enhanced input modalities or hybrid learning strategies.

Future work can explore several directions. One is the integration of structural descriptors derived from atomic coordinates, bond angle distributions, simulated pair distribution functions (PDFs), or symmetry functions like SOAP. These can capture subtle symmetry-breaking effects not visible in XRD peak profiles. Another avenue is the development of multi-modal machine learning models that combine XRD patterns with chemical composition, elemental embeddings, or textual descriptors (e.g., generated by robocrystallographer). This would provide the model with additional context for disambiguating overlapping classes. In addition, the use of 2D XRD formats such as azimuthally resolved diffraction or reciprocal space maps—when available—can preserve angular anisotropy lost in powder averaging, thereby enhancing symmetry resolution.

Beyond the input space, model architecture itself presents an opportunity for advancement. Transfer learning on large-scale pre-trained XRD models followed by fine-tuning on perovskite-specific datasets may significantly improve generalization, particularly for minority symmetry classes. Finally, broader feature engineering strategies such as wavelet and Fourier transforms, although tested in preliminary form here, deserve further systematic evaluation—possibly in combination with deep learning pipelines—to enhance model robustness and interpretability. These extensions will be the focus of future work as we aim to push toward more generalized and accurate symmetry classification systems that remain robust to noise, class imbalance, and real-world data limitations.
}

\section{Results and Discussions}\label{sec3}
Different machine learning models such as Time Series Forest, Random Forest, XGBoost, and neural networks were developed. Tables 1, 2, and 3 show the metrics obtained for the ML models with all possible combinations of augmentations. We preferred macro metrics when evaluating the classifier models on imbalanced datasets due to their ability to provide unbiased evaluation of performance across all classes, regardless of class distribution. Macro metrics calculates metrics such as precision, recall independently and average them, which make sure that performance of minority class is not overshadowed by the majority class. In contrast, weighted class considers the proportion of each class, often leading to the results skewed to majority classes, which leads to poor performance of minority classes. As by using macro metrics, we can gain a clearer understanding of a model’s ability to generalize across all metrics, we selected macro metrics for selecting the best model. However, when reporting the final metrics both the weighted metrics and macro metrics were included to provide a comprehensive understanding of the performance. Weighted metrics reflect the real world scenario providing an understanding of model's overall practical applicability. 

\begin{table}[!h]
\caption{Performance comparison of machine learning algorithms with different data augmentation techniques for crystal system prediction}
\centering
\begin{adjustbox}{width=\textwidth,center}
\begin{tabular}{lcccccc}
\toprule
\multirow{2}{*}{Algorithm} & \multirow{2}{*}{Metric} & \multicolumn{5}{c}{Method} \\
\cmidrule{3-7}
& & Naive & Weighted Class & Jittering & SMOTE & Weighted Class + Jittering \\
\midrule
\multirow{5}{*}{TF} & Precision & \textbf{ 0.9} & 0.85 & 0.89 & 0.84 & 0.85 \\
& Recall & 0.74 & 0.76 & 0.72 & \textbf{0.79} & 0.76 \\
& F1 Score & 0.79 & 0.79 & 0.78 & \textbf{0.81} & 0.79 \\
& MCC & 0.84 & 0.85 & 0.84 & \textbf{0.86} & 0.84 \\
& Accuracy & 90.77 & 90.64 & 90.65 & \textbf{91.77} & 90.64 \\
\midrule
\multirow{5}{*}{RF} & Precision & 0.82 & 0.82 & 0.84 & 0.79 & 0.8 \\
& Recall & 0.6 & 0.64 & 0.55 & 0.74 & 0.56 \\
& F1 Score & 0.65 & 0.69 & 0.6 & 0.76 & 0.61 \\
& MCC & 0.77 & 0.76 & 0.71 & 0.8 & 0.72 \\
& Accuracy & 86.41 & 86.4 & 83.54 & 88.4 & 83.54 \\
\midrule
\multirow{5}{*}{Xgboost} & Precision & 0.8 & 0.76 & 0.72 & 0.79 & 0.67 \\
& Recall & 0.66 & 0.68 & 0.58 & 0.74 & 0.58 \\
& F1 Score & 0.7 & 0.71 & 0.62 & 0.76 & 0.61 \\
& MCC & 0.8 & 0.8 & 0.74 & 0.83 & 0.73 \\
& Accuracy & 88.16 & 88.28 & 85.04 & 89.78 & 84.54 \\
\midrule
\multirow{5}{*}{Neural Networks} & Precision & 0.78 & 0.78 & 0.84 & 0.79 & 0.78 \\
& Recall & 0.74 & 0.76 & 0.77 & 0.77 & 0.76 \\
& F1 Score & 0.75 & 0.76 & 0.79 & 0.77 & 0.77 \\
& MCC & 0.82 & 0.83 & 0.84 & 0.81 & 0.83 \\
& Accuracy & 89.28 & 90.02 & 91.02 & 90.03 & 89.78 \\
\midrule
{\multirow{5}{*}{RNN}} 
& {Precision} & {0.79} & {0.67} & {0.75} & {0.74} & {0.75} \\
& {Recall}    & {0.68} & {0.71} & {0.72} & {0.72} & {0.75} \\
& {F1 Score}  & {0.72} & {0.68} & {0.73} & {0.72} & {0.75} \\
& {MCC}       & {0.79} & {0.76} & {0.80} & {0.81} & {0.82} \\
& {Accuracy}  & {87.78} & {85.29} & {88.40} & {88.65} & {89.40} \\
\midrule
{\multirow{5}{*}{LSTM}} 
& {Precision} & {0.76} & {0.70} & {0.75} & {0.76} & {0.76} \\
& {Recall}    & {0.70} & {0.74} & {0.73} & {0.74} & {0.75} \\
& {F1 Score}  & {0.73} & {0.72} & {0.74} & {0.74} & {0.75} \\
& {MCC}       & {0.81} & {0.80} & {0.81} & {0.82} & {0.82} \\
& {Accuracy}  & {88.65} & {87.91} & {88.90} & {89.15} & {89.40} \\
\midrule
{\multirow{5}{*}{GRU}} 
& {Precision} & {0.79} & {0.69} & {0.78} & {0.75} & {0.76} \\
& {Recall}    & {0.69} & {0.74} & {0.70} & {0.71} & {0.73} \\
& {F1 Score}  & {0.73} & {0.71} & {0.73} & {0.72} & {0.74} \\
& {MCC}       & {0.80} & {0.79} & {0.80} & {0.80} & {0.80} \\
& {Accuracy}  & {87.91} & {87.28} & {88.28} & {88.28} & {88.28} \\
\bottomrule
\end{tabular}
\end{adjustbox}
\label{tab: metrics crystal system}
\end{table}

Table  \ref{tab: metrics crystal system}  shows the performance metrics obtained using ML like time series forest, random forest, XGBoost and neural networks with different augmentation techniques such as jittering, SMOTE,and weighted class. Comparing all metrics, Time series forest with SMOTE is the best-performing model for the XRD crystal system prediction. It showed consistent superior performance across all performance metrics. It achieved a high precision of 0.84, indicating its ability to minimize false positives effectively. The model achieved a recall of 0.79, indicating that the model could identify a substantial proportion of true positives. The F1 score, 0.81 shows a superior performance by the model for the imbalanced dataset. MCC 0.86 demonstrates the robustness of the model to manage the imbalanced dataset. Also, the accuracy of 91.77\% reinforces the reliability of the model. Although methods like random forests and neural networks perform well, they could not consistently surpass time series forest. Therefore, we selected the time series forest with SMOTE augmentation as the best model. The TSF model with different data augmentation strategies outperformed the deep learning approaches. This outcome can be attributed to several factors. First, although we augmented the dataset, its size remains relatively small for deep learning architectures to generalize effectively, whereas classical models are more data-efficient and can generalize well in limited data regimes. Second, the XRD data was standardized to a fixed length of 90, making it moderately low-dimensional. Models like RF and TF are well-suited for such low-dimensional data and tend to perform effectively without the complexity required by deep learning models. \color{black}

\begin{figure}[htbp]
    \centering
    \includegraphics[height=0.9\textheight]{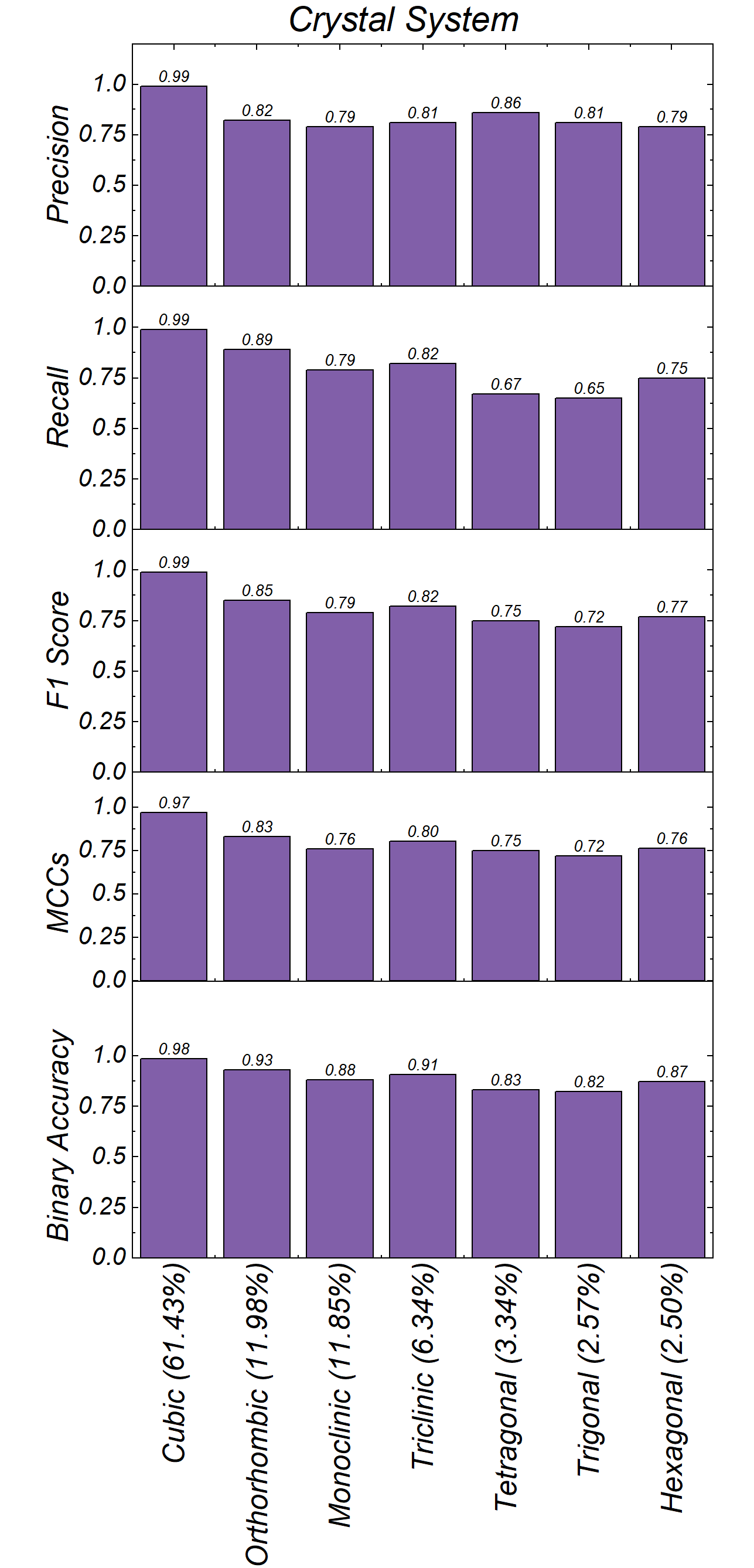} 
    \caption{Performance metrics of the classification model for crystal system prediction using XRD data. Metrics include Precision, Recall, F1 Score, Matthews Correlation Coefficient (MCC), and Binary Accuracy for each crystal system. {X-axis classes are sorted by sample ratio, with percentages shown in each label.}}
    \label{fig:crystal_system}
\end{figure}

Figure \ref{fig:crystal_system} shows the performance metrics of the classification model for predicting crystal systems using XRD data of perovskite materials.  It represents the class-wise metrics for the best model selected. As the data is highly imbalanced, it is important to analyze these metrics for each class. All individual class metrics are greater than or equal to 0.75, except for recall, where the lowest value is around 0.6. This is acceptable when considering the class distribution depicted in Figure 2. We observe classes comprising only 2.5\% and 2.6\% of the whole dataset. The cubic system demonstrates the highest performance across all metrics, with precision, recall, and F1 score values around 0.99, an MCC of 0.97, and a binary accuracy of 0.98. This indicates that the model effectively identifies cubic structures, likely due to their distinct and symmetric XRD patterns. The orthorhombic and monoclinic systems achieve balanced performance, with F1 scores of 0.85 and 0.79, respectively, along with strong MCC and binary accuracy values. In comparison, the trigonal and tetragonal systems exhibit relatively lower performance. The recall for the trigonal system is the lowest at 0.65, indicating a higher rate of false negatives. The F1 score (0.72) and MCC (0.72) also reflect this trend, suggesting difficulties in distinguishing trigonal patterns from other crystal systems. This could be attributed to the overlapping diffraction features with other classes.  The hexagonal system shows moderate performance, with recall (0.75) slightly lower than precision (0.79), suggesting that the model is more conservative in predicting hexagonal structures. 
{
To further examine the effect of class imbalance, Figure~\ref{fig:crystal_system} has been updated to display class-wise performance metrics sorted by descending data ratio, with percentage labels included on the x-axis. A clear positive correlation is observed between class frequency and prediction accuracy. The cubic system, which constitutes the largest portion of the dataset, achieves near-perfect F1, recall, and MCC values. In contrast, low-frequency classes such as trigonal and tetragonal exhibit noticeably lower performance, particularly in recall. These results validate the importance of class-balancing techniques such as SMOTE and jittering, which help improve model sensitivity for underrepresented crystal systems.
}

\begin{figure}[htbp]
    \centering
    \includegraphics[height=0.9\textheight]{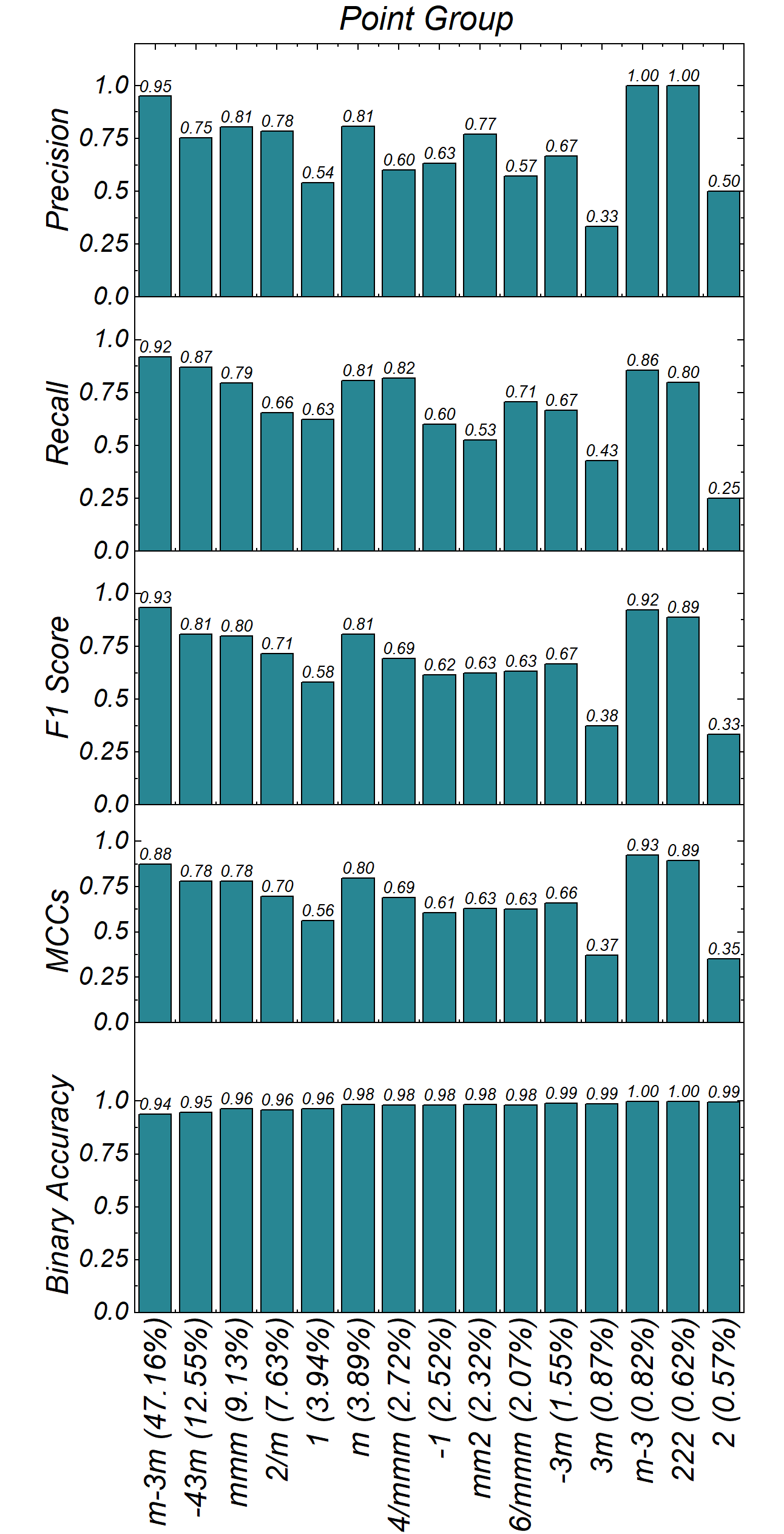} 
    \caption{Performance metrics of the classification model for point group prediction using XRD data. Metrics include Precision, Recall, F1 Score, Matthews Correlation Coefficient (MCC), and Binary Accuracy for each point group: $-1$, $-3m$, $-43m$, $1$, $2$, $2/m$, $222$, $3m$, $4/mmm$, $6/mmm$, $m$, $m-3$, $m-3m$, and $mmm$.{X-axis classes are sorted by sample ratio, with percentages shown in each label.}}    \label{fig:point_group}
\end{figure}

\begin{table}[htbp]
\caption{Performance comparison of machine learning algorithms with different data augmentation techniques for point group prediction}
\centering
\begin{adjustbox}{width=\textwidth,center}
\begin{tabular}{lcccccc}
\toprule
\multirow{2}{*}{Algorithm} & \multirow{2}{*}{Metric} & \multicolumn{5}{c}{Method} \\
\cmidrule{3-7}
& & Naive & Weighted Class & Jittering & SMOTE & Weighted Class + Jittering \\
\midrule
\multirow{5}{*}{TF} 
& Precision & 0.69 & 0.68 & 0.71 & 0.71 & 0.70 \\
& Recall & 0.62 & 0.59 & 0.59 & \textbf{0.69} & 0.59 \\
& F1 Score & 0.65 & 0.62 & 0.63 & \textbf{0.69} & 0.63 \\
& MCC & \textbf{0.76} & 0.74 & 0.74 & 0.75 & 0.75 \\
& Accuracy & \textbf{82.51} & 81.11 & 81.12 & 81.12 & 81.62 \\
\midrule
\multirow{5}{*}{RF} 
& Precision & \textbf{0.81} & 0.73 & 0.80 & 0.72 & 0.72 \\
& Recall & 0.58 & 0.57 & 0.47 & 0.64 & 0.45 \\
& F1 Score & 0.65 & 0.62 & 0.53 & 0.66 & 0.50 \\
& MCC & 0.72 & 0.72 & 0.67 & 0.75 & 0.65 \\
& Accuracy & 80.1 & 79.47 & 76.17 & 81.11 & 74.65 \\
\midrule
\multirow{5}{*}{Xgboost} 
& Precision & 0.64 & 0.68 & 0.59 & 0.62 & 0.60 \\
& Recall & 0.53 & 0.58 & 0.46 & 0.60 & 0.49 \\
& F1 Score & 0.56 & 0.61 & 0.49 & 0.60 & 0.53 \\
& MCC & 0.70 & 0.71 & 0.62 & 0.72 & 0.66 \\
& Accuracy & 78.2 & 78.83 & 74.02 & 79.47 & 75.67 \\
\midrule
\multirow{5}{*}{Neural Networks} 
& Precision & 0.63 & 0.55 & 0.62 & 0.64 & 0.64 \\
& Recall & 0.55 & 0.55 & 0.59 & 0.63 & 0.61 \\
& F1 Score & 0.57 & 0.54 & 0.59 & 0.63 & 0.61 \\
& MCC & 0.67 & 0.66 & 0.65 & 0.69 & 0.68 \\
& Accuracy & 74.68 & 73.92 & 74.16 & 75.7 & 75.94 \\

\midrule
{\multirow{5}{*}{RNN}} 
& {Precision} & {0.62} & {0.57} & {0.59} & {0.60} & {0.57} \\
& {Recall}    & {0.55} & {0.61} & {0.58} & {0.60} & {0.59} \\
& {F1 Score}  & {0.57} & {0.58} & {0.58} & {0.59} & {0.57} \\
& {MCC}       & {0.67} & {0.64} & {0.69} & {0.66} & {0.66} \\
& {Accuracy}  & {75.92} & {71.99} & {76.93} & {73.89} & {73.64} \\
\midrule
{\multirow{5}{*}{LSTM}} 
& {Precision} & {0.60} & {0.61} & {0.66} & {0.60} & {0.56} \\
& {Recall}    & {0.57} & {0.60} & {0.62} & {0.62} & {0.61} \\
& {F1 Score}  & {0.58} & {0.59} & {0.64} & {0.60} & {0.57} \\
& {MCC}       & {0.68} & {0.66} & {0.73} & {0.66} & {0.67} \\
& {Accuracy}  & {76.17} & {73.38} & {80.48} & {73.64} & {74.78} \\
\midrule
{\multirow{5}{*}{GRU}} 
& {Precision} & {0.63} & {0.59} & {0.62} & {0.65} & {0.62} \\
& {Recall}    & {0.59} & {0.59} & {0.58} & {0.64} & {0.63} \\
& {F1 Score}  & {0.60} & {0.57} & {0.59} & {0.63} & {0.61} \\
& {MCC}       & {0.69} & {0.64} & {0.69} & {0.67} & {0.67} \\
& {Accuracy}  & {77.06} & {72.37} & {77.44} & {74.78} & {74.27} \\

\bottomrule
\end{tabular}
\end{adjustbox}
\label{tab:metrics point group}
\end{table}

Table  \ref{tab:metrics point group}  shows the metrics obtained for the XRD classification for point prediction. We considered 15 classes of point groups for the classification, where the data is highly imbalanced. More than 75\% of the dataset belongs to four classes, while the remaining 25\% is distributed among other 11 classes. As the data set is highly imbalanced, we focus on metrics such as MCC and F1 score to select the model. From the table, we observe that the TSF model with SMOTE augmentation performs consistently well across all metrics, even though it could not provide the best individual scores. The TSF model with SMOTE is able to achieve an F1 score of 0.69 and a recall of 0.61. The precision of the random forest model is the highest, with a value of 0.81, but its performance on all other metrics is poor. The naive TSF model achieved an MCC of 0.76, whereas the SMOTE-based TSF model achieved 0.75. As the TSF model with SMOTE performed comparatively well across all models, we selected it as the best model for point group prediction. {While a general trend of improved performance with higher data representation is observed, notable exceptions emerge. For example, point groups such as $2/m$ and $222$ show poor F1 scores (below 0.4) despite moderate class frequencies, likely due to structural ambiguity and spectral overlap. In contrast, high-symmetry groups like $m\overline{3}m$ and $3m$ demonstrate excellent classification performance across all metrics, reflecting their distinct and well-separated diffraction signatures.
}

\begin{table}[htbp]
\caption{Performance comparison of machine learning algorithms with different data augmentation techniques for space group prediction}
\centering
\begin{adjustbox}{width=\textwidth,center}
\begin{tabular}{lcccccc}
\toprule
\multirow{2}{*}{Algorithm} & \multirow{2}{*}{Metric} & \multicolumn{5}{c}{Method} \\
\cmidrule{3-7}
& & Naive & Weighted Class & Jittering & SMOTE & Weighted Class + Jittering \\
\midrule
\multirow{5}{*}{TF} 
& Precision & 0.81 & 0.66 & 0.79 & 0.75 & \textbf{0.81} \\
& Recall & 0.73 & 0.60 & 0.70 & 0.73 & 0.72 \\
& F1 Score & 0.75 & 0.63 & 0.73 & 0.73 & \textbf{0.75} \\
& MCC & 0.80 & 0.74 & 0.79 & 0.80 & \textbf{0.80} \\
& Accuracy & 83.29 & 81.08 & 83.01 & 83.71 & \textbf{83.83} \\
\midrule
\multirow{5}{*}{RF} 
& Precision & 0.81 & 0.80 & 0.78 & 0.70 & 0.76 \\
& Recall & 0.71 & 0.71 & 0.60 & 0.64 & 0.60 \\
& F1 Score & 0.73 & 0.72 & 0.64 & 0.65 & 0.64 \\
& MCC & 0.81 & 0.79 & 0.75 & 0.74 & 0.74 \\
& Accuracy & 84.27 & 82.58 & 79.76 & 80.42 & 78.93 \\
\midrule
\multirow{5}{*}{Xgboost} 
& Precision & 0.72 & 0.70 & 0.59 & 0.70 & 0.56 \\
& Recall & 0.70 & 0.69 & 0.54 & 0.70 & 0.54 \\
& F1 Score & 0.70 & 0.68 & 0.55 & 0.69 & 0.54 \\
& MCC & 0.77 & 0.76 & 0.73 & 0.77 & 0.71 \\
& Accuracy & 81.46 & 80.76 & 77.81 & 81.18 & 76.54 \\
\midrule
\multirow{5}{*}{Neural Networks} 
& Precision & 0.67 & 0.72 & 0.67 & 0.64 & 0.68 \\
& Recall & 0.67 & 0.67 & 0.65 & 0.63 & 0.67 \\
& F1 Score & 0.65 & 0.66 & 0.64 & 0.64 & 0.64 \\
& MCC & 0.71 & 0.72 & 0.70 & 0.72 & 0.74 \\
& Accuracy & 75.84 & 76.97 & 74.16 & 76.12 & 78.37 \\

\midrule
{\multirow{5}{*}{RNN}} 
& {Precision} & {0.70} & {0.70} & {0.72} & {0.73} & {0.72} \\
& {Recall}    & {0.68} & {0.69} & {0.72} & {0.74} & {0.73} \\
& {F1 Score}  & {0.68} & {0.68} & {0.71} & {0.73} & {0.71} \\
& {MCC}       & {0.74} & {0.74} & {0.77} & {0.76} & {0.75} \\
& {Accuracy}  & {78.51} & {78.65} & {80.90} & {79.78} & {79.92} \\
\midrule
{\multirow{5}{*}{LSTM}} 
& {Precision} & {0.74} & {0.68} & {0.76} & {0.74} & {0.72} \\
& {Recall}    & {0.73} & {0.67} & {\textbf{0.74}} & {0.74} & {0.70} \\
& {F1 Score}  & {0.73} & {0.64} & {0.74} & {0.74} & {0.70} \\
& {MCC}       & {0.75} & {0.72} & {0.77} & {0.76} & {0.76} \\
& {Accuracy}  & {79.63} & {77.25} & {81.04} & {79.63} & {80.20} \\
\midrule
{\multirow{5}{*}{GRU}} 
& {Precision} & {0.72} & {0.70} & {0.73} & {0.74} & {0.72} \\
& {Recall}    & {0.68} & {0.69} & {0.73} & {0.75} & {0.72} \\
& {F1 Score}  & {0.70} & {0.68} & {0.72} & {0.74} & {0.70} \\
& {MCC}       & {0.75} & {0.74} & {0.77} & {0.76} & {0.76} \\
& {Accuracy}  & {79.21} & {78.65} & {81.18} & {79.92} & {80.62} \\

\bottomrule
\end{tabular}
\end{adjustbox}

\label{tab:metrics space group}
\end{table}
Figure \ref{fig:point_group} shows the performance metrics of the classification model for predicting point groups. Model demonstrates strong performance for certain point groups such as $3m$, $m-3$, and $m-3m$, where precision, recall, F1 score, and MCC are all above 0.9, with binary accuracy reaching 1.00 for $3m$ and $m-3$. This indicates high confidence and accuracy in classifying these groups, likely due to their distinct symmetry-related diffraction features. Conversely, some point groups, such as $2/m$ and $222$, show lower performance. For $2/m$, the precision is 0.33, recall is 0.43, and the F1 score is 0.38, indicating difficulties in both correctly identifying and retrieving true positive samples. Similarly, $222$ has a low recall of 0.25 and an F1 score of 0.33, suggesting significant misclassification issues, which may arise from overlapping diffraction patterns with other point groups or insufficient data representation for these categories. Intermediate performance is observed in point groups like $-1$, $-3m$, and $6/mmm$, where precision, recall, and F1 scores are moderately high, ranging from 0.6 to 0.8. The binary accuracy remains consistently high across all point groups, exceeding 0.94, reflecting the model's overall ability to distinguish between positive and negative cases effectively. Overall, the classification model performs well for point groups with clear symmetry features while facing challenges with groups that have complex or less distinct diffraction characteristics.

\begin{figure}[htbp]
    \centering
    \includegraphics[height=0.9\textheight]{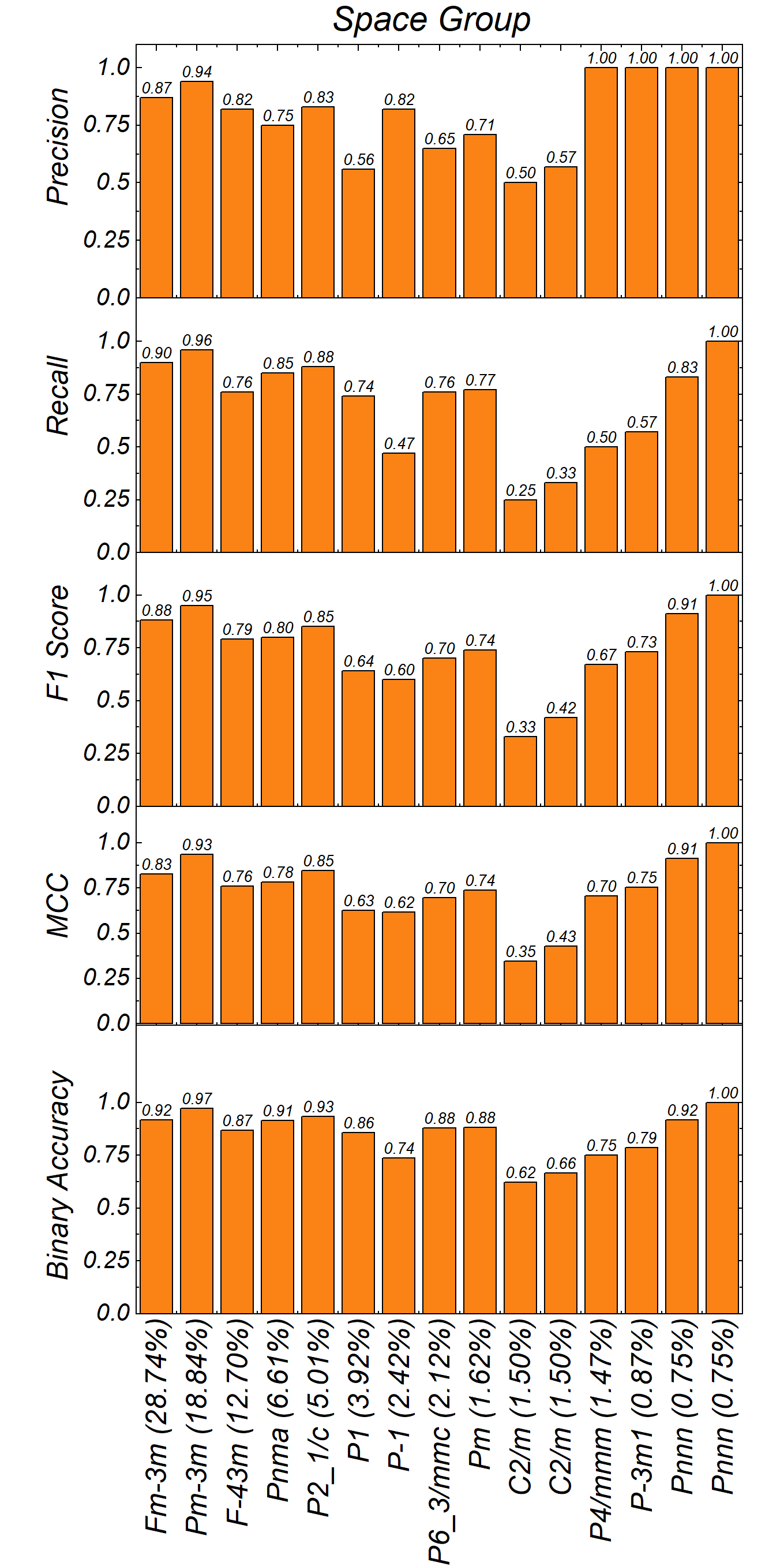} 
    \caption{Performance metrics of the classification model for space group prediction : $C2/m$, $Cm$, $F-43m$, $Fm-3m$, $P-1$, $P-3m1$, $P1$, $P2_1/c$, $P4/mmm$, $P6_3/mmc$, $Pc$, $Pm$, $Pm-3m$, $Pnma$, and $Pnnn$. {X-axis classes are sorted by sample ratio, with percentages shown in each label.}}
    \label{fig:space_group}
\end{figure}

{
To further interpret the performance trends observed in Figure 5 and Figure 6, we also explored the intrinsic structure of the XRD-based chemical space using unsupervised clustering. Principal Component Analysis (PCA) was applied to the 90-point interpolated XRD vectors, followed by KMeans clustering to probe the separability of symmetry classes without label supervision. While high-symmetry point groups such as $m\overline{3}m$ and $\overline{4}3m$ showed some cluster formation, low-symmetry groups like $2/m$, $222$, and $\overline{1}$ were substantially overlapped in latent space. These findings mirror the supervised classification results, where low-symmetry classes exhibited lower accuracy and recall. This alignment suggests that symmetry-specific features are either subtle or nonlinear and may not manifest clearly in spectral intensity patterns alone. Consequently, future efforts may benefit from hybrid models that integrate XRD data with complementary structural descriptors to improve class separability and predictive robustness.
}

\begin{table}[htbp]
\caption{Performance Metrics for the best models selected for crystal system. point group and apace group prediction}
\centering
\begin{tabularx}{\textwidth}{l|X|X|X}
\hline
\textbf{Target} & \textbf{Crystal system } & \textbf{Point group } & \textbf{Space group } \\ \hline
\textbf{Model Selected} & TSF SMOTE & TSF SMOTE & TSF weighted class + jittering \\ \hline
Precision (Macro avg)            & 0.84  & 0.71  & 0.81  \\ 
Precision (Weighted avg)         & 0.92  & 0.83  & 0.84  \\ \hline
Recall (Macro avg)               & 0.79  & 0.69  & 0.72  \\ 
Recall (Weighted avg)            & 0.92  & 0.0.83  & 0.84  \\ \hline
F1 Score (Macro avg)             & 0.81  & 0.69  & 0.75  \\ 
F1 Score (Weighted avg)          & 0.92  & 0.83  & 0.84  \\ \hline
MCC (Macro avg)                  & 0.86  & 0.75  & 0.8    \\ 
MCC (Weighted avg)               & 0.9 & 0.79 & 0.8 \\ \hline
Accuracy              & 91.77 & 81.12    & 83.83 \\ 
Binary accuracy (Weighted avg) & 97.76 & 95.27 & 95.18 \\ \hline
\end{tabularx}
\label{tab: macro vs weighted}
\end{table}

For the space group prediction, as shown in table \ref{tab:metrics space group} the TSF model with weighted class and jittering showed the most robust and consistent results across all metrics, specifically Precision (0.81), F1 Score (0.75), MCC (0.80), and Accuracy (83.83\%). The LSTM models achieved a recall of 0.74, which indicates that while these techniques improved data balance, they did not significantly enhance model sensitivity (the ability to predict true positives)\color{black}. On the other hand, the TSF model with the weighted class method provided balanced performance across all metrics. Therefore, we selected the TSF model with jittering and weighted class augmentation as the best technique for space group prediction. {
Figure ~\ref{fig:space_group} shows the performance metrics of the classification model for predicting space groups using XRD data of perovskite materials. The model demonstrates excellent classification performance for space groups such as $Pm\overline{3}m$ and $Pnnn$, with precision, recall, F1 score, and MCC all approaching or reaching 1.00. Notably, $Pnnn$ achieves these results despite comprising only 0.75\% of the dataset, while $Pm\overline{3}m$ achieves an F1 score of 0.95 and MCC of 0.93. In contrast, $Pnma$, although well-represented with a data ratio of 6.61\%, shows comparatively lower performance (F1 = 0.80, MCC = 0.78). This indicates solid but not near-perfect classification, likely due to spectral overlap with other orthorhombic space groups or internal structural variability. These results highlight that strong model performance depends not only on data balance but also on the distinctiveness of spectral features within each class.
}.In contrast, the space groups $C2/m$ and $Cm$ show relatively lower performance. For $C2/m$, precision and recall are 0.50 and 0.25, respectively, leading to an F1 score of 0.33 and an MCC of 0.35. This indicates challenges in correctly classifying samples belonging to this space group, possibly due to overlapping diffraction peaks with other groups or insufficient data representation. Similarly, $Cm$ shows a recall of 0.33 and an F1 score of 0.42, suggesting classification difficulties. Intermediate performance is observed for space groups such as $P1$, $P-3m1$, and $P4/mmm$, with F1 scores ranging from 0.64 to 0.85 and MCC values between 0.63 and 0.85. These results suggest a moderate level of classification accuracy, likely influenced by the complexity of their diffraction patterns. The binary accuracy remains consistently high across most space groups, exceeding 0.74, reflecting the model's robustness in distinguishing between positive and negative classifications. However, for lower-performing groups like $C2/m$, binary accuracy drops to 0.62, consistent with the other performance metrics. {While some high-data classes like Pnnn and Pm$\overline{3}$m achieve excellent results, others like Pnma perform below expectation. Low-performing groups such as C2/m and Cm suggest that spectral overlap or internal variability, not just data imbalance, limits accuracy. This underscores the need for both balanced data and spectrally distinct features in fine-grained symmetry classification.
} Table \ref{tab: macro vs weighted} presents the results of the final model selected for crystal system, point group, and space group prediction. Since weighted metrics account for class distribution, we observe that they perform significantly better than macro metrics. This validates the applicability of our model in real-world scenarios.The weighted MCC for the crystal system is 0.9, for the point group is 0.79, and for the space group is 0.8. The binary accuracy for the crystal system is 97.76\%, for the point group is 95.27\%, and for the space group is 95.18\%.The Precision, recall, and F1 score also demonstrate improved values for the weighted metrics.{
Compared to existing literature, these results are highly competitive. For instance, Kaufmann et al.~\cite{kaufmann2020crystal} reported $\sim$93.5\% accuracy for Bravais lattice prediction using ResNet50 on electron diffraction data, while Chen et al.~\cite{Chen2024} achieved strong space group classification using a ResNet trained on over 60,000 XRD patterns. Salgado et al.~\cite{Salgado2023} obtained $\sim$86\% accuracy for both crystal system and space group tasks using deep learning models trained on ICSD data, and Oviedo et al.~\cite{Oviedo2019} reached 89\% space group accuracy on small CNN-augmented datasets. In contrast, our TSF-based models operate directly on raw interpolated XRD spectra with minimal preprocessing and no hand-crafted features. They generalize effectively under data-scarce and class-imbalanced conditions—challenges that often impair CNN and transformer-based models. Furthermore, our workflow integrates diffraction-specific augmentation strategies such as jittering and SMOTE for spectral sequences and evaluates alternative input representations (e.g., wavelet and Fourier transforms), offering insights rarely explored in prior studies. Importantly, the pipeline is built entirely on DFT-relaxed structures from the Materials Project, enabling reproducibility and scalability.
}

{
Perovskite materials exhibit a wide range of chemical compositions that influence their crystallographic symmetry and, consequently, the interpretability of their XRD patterns. To investigate this effect, we categorized the 4,009 perovskite entries into four composition classes based on the presence of anionic species: Oxide, Halide, Mixed (Halide + Oxide), and Other (Chalcogenides, Nitrides, Pnictides, etc.). This classification was determined by parsing chemical formulas and identifying key anionic groups such as O$^{2-}$, F$^{-}$, Cl$^{-}$, Br$^{-}$, I$^{-}$, S$^{2-}$, N$^{3-}$, and P$^{3-}$. The oxide group dominates the dataset (2,702 entries), followed by halides (945), with mixed and other classes contributing 25 and 336 entries, respectively.

Figure~\ref{fig:composition} provides a breakdown of these composition types across symmetry classes, revealing distinct compositional biases. Halide perovskites predominantly crystallize in highly symmetric structures such as cubic ($Fm\overline{3}m$, $Pm\overline{3}m$) and are heavily concentrated in point groups like $m\overline{3}m$ and $\overline{4}3m$. Their diffraction patterns tend to be sharp and well-separated, resulting in clearer symmetry labeling and enhanced model performance. In contrast, oxide perovskites span a broader symmetry landscape, including monoclinic, orthorhombic, and triclinic systems (e.g., $P2_1/c$, $Pnma$, $P\overline{1}$). These structures often exhibit peak broadening and overlap due to octahedral tilting and distortions driven by ionic radii mismatch and Jahn–Teller effects, making classification more challenging. Other compositions, such as chalcogenides and pnictides, mainly occur in lower-symmetry groups and contribute to spectral diversity. While they offer generalization potential, their low frequency and non-standard bonding environments often lead to misclassification. Mixed halide–oxide perovskites are sparsely represented ($n=25$) and lack a dominant structural class, reflecting their compositional complexity and limited availability in open databases. These compositional effects correlate strongly with classification outcomes. Halide-dominated symmetry classes achieve MCC values above 0.9, while oxide-rich classes often fall below 0.75. These discrepancies cannot be fully attributed to class imbalance; they stem from intrinsic spectral ambiguity driven by chemical complexity. This confirms that symmetry classification from XRD is not solely a geometric task but is closely linked to composition. We also note that the “Other” category (including S-, N-, and P-based compounds) contributes useful structural diversity but is currently underutilized due to low representation. Future work could improve robustness by incorporating domain-informed features such as ionic radius ratios and electronegativity, or by developing hierarchical classifiers tailored to specific compositional regimes.
}

\section{{Conclusion}}\label{sec:4}

{
This work presents a machine learning framework for classifying the crystal system, point group, and space group of perovskite materials using powder XRD patterns. A filtered dataset of 4,009 DFT-relaxed structures from the Materials Project was compiled to ensure thermodynamic relevance and label clarity. To address class imbalance and spectral complexity, we applied SMOTE, jittering, and class weighting.Among several evaluated models, Time Series Forest (TSF) consistently outperformed others by treating XRD spectra as sequential signals. It achieved 97.76\% accuracy and 0.90 MCC for crystal system prediction. For point group classification, TSF reached 95.27\% balanced accuracy and 0.79 MCC, while for space group prediction, it achieved 95.18\% balanced accuracy and 0.80 MCC. The model was especially reliable for high-symmetry classes with distinctive diffraction patterns, while challenges remained for overlapping or low-symmetry cases. We also analyzed compositional trends, classifying entries into oxides (67\%), halides (24\%), and mixed or other types. These groups showed distinct symmetry distributions, highlighting a strong correlation between anionic chemistry and structural complexity. Although trained on simulated patterns, our models provide a reproducible and interpretable foundation for real-world application. Future work includes benchmarking on experimental XRD datasets to assess robustness under conditions such as texturing and peak loss. Overall, this study introduces a scalable and data-efficient approach for symmetry classification from XRD, offering insights into both structural patterns and compositional influences, and supporting high-throughput materials discovery workflows.
}

\section{{Data and Code Availability}}\label{sec:5}

{
All structural data used in this study were obtained from the publicly accessible Materials Project (MP) database via the Pymatgen API. The specific Material IDs, XRD patterns, and symmetry labels used for model training can be regenerated using the filtering criteria outlined in Section~2.1.

The codebase used for model development is currently under institutional confidentiality, as it was developed within a government research setting (Dubai Electricity and Water Authority, DEWA). Access can be requested by contacting the corresponding author or DEWA R\&D directly via official channels. We are committed to supporting academic and collaborative transparency, and will provide access for non-commercial use subject to institutional approval.}

\section{Acknowledgements}\label{sec5}
Authors acknowledge the support provided by the Dubai Electricity and Water Authority (DEWA)'s Research and Development Center for this study.

 \bibliographystyle{elsarticle-num} 
 \bibliography{cas-refs}
\appendix
\renewcommand{\thefigure}{A\arabic{figure}}  
\setcounter{figure}{0}  
\section{Appendix Section}
\label{sec:sample:appendix}
Below are the tables comparing the time series forest model with different normalization and augmentation techniques. The naive TSF model serves as the baseline for comparison. Table \ref{tab:row-norm} shows the metrics for row-wise normalized data as inputs. Table \ref{tab:std-norm} presents the metrics when the data is normalized using a standard scaler. The metrics are notably low for these two preprocessing methods. Table \ref{tab:scaling} displays the performance metrics when the data is augmented, while Table \ref{tab:shifting} shows the metrics when the data is augmented using shifting. These two augmentations did not positively impact the metrics. Finally, Table \ref{tab:Features} presents the metrics obtained after converting the data into the features discussed in section \ref{sec2}. The extracted features did not perform better than the naive model.

\begin{table*}[H]
\centering
\begin{minipage}[b]{0.45\linewidth}
\centering
\caption{Performance Metrics Comparison of TF Algorithm (Naive vs Row-Wise Normalisation)}
\begin{adjustbox}{width=\textwidth, center}
\begin{tabular}{@{}lcc@{}}
\toprule
\textbf{Metric} & \textbf{Naive} & \textbf{Row-Wise Norm} \\ \midrule
Precision & 90 & 0.76 \\
Recall    & 74 & 0.64 \\
F1 Score  & 79 & 0.67 \\
MCC       & 0.84 & 0.78 \\
Accuracy  & 90.77 & 87.16 \\ \bottomrule
\end{tabular}

\end{adjustbox}
\label{tab:row-norm}
\end{minipage}
\hfill
\begin{minipage}[b]{0.45\linewidth}
\centering
\caption{Performance Metrics Comparison of TF Algorithm (Naive vs Standard Scaler)}
\begin{adjustbox}{width=\textwidth, center}
\begin{tabular}{@{}lcc@{}}
\toprule
\textbf{Metric} & \textbf{Naive} & \textbf{Standard Scaler} \\ \midrule
Precision & 90 & 0.76 \\
Recall    & 74 & 0.63 \\
F1 Score  & 79 & 0.67 \\
MCC       & 0.84 & 0.75 \\
Accuracy  & 90.77 & 86.16 \\ \bottomrule
\end{tabular}

\end{adjustbox}
\label{tab:std-norm}
\end{minipage}

\vspace{1em}

\begin{minipage}[b]{0.45\linewidth}
\centering
\caption{Performance Metrics Comparison of TF Algorithm (Naive vs Shifting Aug)}
\begin{adjustbox}{width=\textwidth, center}
\begin{tabular}{@{}lcc@{}}
\toprule
\textbf{Metric} & \textbf{Naive} & \textbf{Shifting Aug} \\ \midrule
Precision & 90 & 0.17 \\
Recall    & 74 & 0.15 \\
F1 Score  & 79 & 0.12 \\
MCC       & 0.84 & 0.07 \\
Accuracy  & 90.77 & 61.72 \\ \bottomrule
\end{tabular}

\end{adjustbox}
\label{tab:shifting}
\end{minipage}
\hfill
\begin{minipage}[b]{0.45\linewidth}
\centering
\caption{Performance Metrics Comparison of TF Algorithm (Naive vs Scaling Aug)}
\begin{adjustbox}{width=\textwidth, center}
\begin{tabular}{@{}lcc@{}}
\toprule
\textbf{Metric} & \textbf{Naive} & \textbf{Scaling Aug} \\ \midrule
Precision & 90 & 0.19 \\
Recall    & 74 & 0.16 \\
F1 Score  & 79 & 0.14 \\
MCC       & 0.84 & 0.11 \\
Accuracy  & 90.77 & 59.85 \\ \bottomrule
\end{tabular}

\end{adjustbox}
\label{tab:scaling}
\end{minipage}

\vspace{1em}

\begin{minipage}[b]{0.45\linewidth}
\centering
\caption{Performance Metrics Comparison of TF Algorithm (Naive vs Features Naïve)}
\begin{adjustbox}{width=\textwidth, center}
\begin{tabular}{@{}lcc@{}}
\toprule
\textbf{Metric} & \textbf{Naive} & \textbf{Features Naïve} \\ \midrule
Precision & 90 & 0.67 \\
Recall    & 74 & 0.57 \\
F1 Score  & 79 & 0.60 \\
MCC       & 0.84 & 0.70 \\
Accuracy  & 90.77 & 82.92 \\ \bottomrule
\end{tabular}

\end{adjustbox}
\label{tab:Features}
\end{minipage}
\end{table*}

Table A.10 presents a performance comparison between raw XRD data and various transformed representations, including Fourier Transform (magnitude only and magnitude + phase) and Wavelet Transform. The model trained on raw data outperformed all others across all metrics, achieving the highest precision (0.90), F1 score (0.79), MCC (0.84), and accuracy (90.77\%). While the Wavelet representation showed a more balanced trade-off between precision and recall, the overall performance of transformed representations remained lower than the raw input.
\begin{table}[ht]
\centering
\caption{{Performance comparison between raw XRD data and transformed representations}}
\begin{tabular}{lcccc}
\hline
{\textbf{Metric}} & {\textbf{Naïve (Raw)}} & {\textbf{FT (Magnitude)}} & {\textbf{FT (Mag + Phase)}} & {\textbf{Wavelet}} \\
\hline
{Precision} & {0.90} & {0.81} & {0.88} & {0.87} \\
{Recall}    & {0.74} & {0.60} & {0.63} & {0.66} \\
{F1 Score}  & {0.79} & {0.64} & {0.68} & {0.71} \\
{MCC}       & {0.84} & {0.78} & {0.80} & {0.889} \\
{Accuracy (\%)} & {90.77} & {87.28} & {88.27} & {81.00} \\
\hline
\end{tabular}
\label{tab:transformed_comparison}
\end{table}

\color{black}

\subsection{ Information  on Optimised Hyperparameters for best models}
The following configurations are the hyperparameters obtained for our best model using grid search cross-validation:

\begin{itemize}
    \item \textbf{Crystal System Prediction (TSF + SMOTE):}
    \begin{itemize}
        \item \texttt{n\_estimators: 100}
        \item \texttt{max\_depth: 20}
        \item \texttt{min\_samples\_split: 2}
        \item \texttt{min\_samples\_leaf: 1}
    \end{itemize}
    
    \item \textbf{Symbol Prediction (TSF + SMOTE):}
    \begin{itemize}
        \item \texttt{n\_estimators: 200}
        \item \texttt{max\_depth: 20}
        \item \texttt{min\_samples\_split: 2}
        \item \texttt{min\_samples\_leaf: 1}
    \end{itemize}
    
    \item \textbf{Point Group Prediction (TSF + Weighted Class + Jittering):}
    \begin{itemize}
        \item \texttt{n\_estimators: 200}
        \item \texttt{max\_depth: 30}
        \item \texttt{min\_samples\_split: 2}
        \item \texttt{min\_samples\_leaf: 1}
    \end{itemize}
\end{itemize}

Below, we provide a brief explanation of the models we used. 
\subsubsection{Random Forest}\label{subsubsec2}
Random forest is a bagging technique where the model is an ensemble of decision trees. Combining decision trees results in more general learning and keeps the model from overfitting. Each tree is grown by randomly selecting its features and samples. The training of the decision trees happens independently of each other. The final output is the average across all decision trees. Random Forest is famous for its robustness, versatility, and ability to handle high-dimensional data. The drawbacks of this ML model are its high computation time and the need for hyperparameter tuning. 
\subsubsection{Time Series Forest}\label{subsubsec2}
Time series forest(TSF)\cite{deng2013timeseriesforestclassification} is an ensemble tree method for time series forecasting and classification. The time series tree is the fundamental component of the TSF. The peculiarities of TSF are interval features and entrance gain. Features of a TSF are formed by dividing the time series into multiple intervals and extracting the features such as mean, standard deviation, and slope corresponding to that interval. TSF randomly samples features at each tree node, which implies that each node evaluates different sets of features. This random sampling of features helps capture various aspects of time series data. The model estimates the splits of trees based on entropy and distance measures (entrance gain). Entrance gain aids in improving the accuracy of the model while predicting for time series. TSF outputs the majority class voted by time series trees. We used a time series forest to analyze XRD data as it is very similar to time series data. Instead of time in x-axis , it is theta which varies from 0-90 degree. 
\subsubsection{Neural Network}\label{subsubsec2}
A neural network is an AI model where the human brain inspires the architecture. The model comprises an input, hidden, and output layer. Neurons are the fundamental unit of the neural network where every neuron is interconnected and associated with weights and activation functions. The activation function introduces non-linearity to the model, which enables the model to learn the complex patterns of the data. The network learns from the data by updating weights and minimizing the loss between the prediction and the actual through the gradient descent algorithm. The model selected for this work has two hidden layers, each consisting of 128 and 64 neurons, respectively. The model we designed is computationally efficient. However, the model may struggle to grasp the detailed patterns in XRD data because it lacks a specialized mechanism to understand the order or sequence of information.
\subsubsection{XGBoost}\label{subsubsec2}

eXtreme Gradient Boosting (XGBoost)\cite{Chen_2016} is a boosting ensemble ML algorithm that utilizes decision trees [15]. It consists of a sequence of decision trees, where each tree learns from the errors of the previous trees. The model is trained by optimizing a loss function, which is a combination of a data loss function and a regularization loss function. XGBoost is well-known among data practitioners for its high accuracy on well-structured data. Additionally, it is robust against overfitting due to its built-in regularization techniques.






\end{document}